# Timescales for Nitric Acid Desorption in Epitaxial Graphene Devices

Swapnil M. Mhatre[1,2], Ngoc Thanh Mai Tran[1,3], Heather M. Hill[1], Dipanjan Saha[1], Angela R. Hight Walker[1], Chi-Te Liang[2], Randolph E. Elmquist[1], David B. Newell[1], and Albert F. Rigosi[1]*

*[1]Physical Measurement Laboratory, National Institute of Standards and Technology (NIST), Gaithersburg, MD 20899, United States*

*[2]Graduate Institute of Applied Physics, National Taiwan University, Taipei 10617, Taiwan*

*[3]Joint Quantum Institute, University of Maryland, College Park, Maryland 20742, United States*

(Received 22 March 2022;)

ABSTRACT: This work reports the dynamics of transient hole doping in epitaxial graphene devices by using nitric acid as an adsorbent. The timescales associated with corresponding desorption processes are extracted from the data. The understanding of reversible hole doping without gating is of crucial importance to those fabricating devices with a particular functionality. Measurements of the electrical and optical properties of several devices post-exposure were performed with transport temperatures between 300 K and 1.5 K. Ambient conditions are applied to non-transport measurements to replicate the most likely laboratory conditions for handling devices using this doping method. The relevant timescales from transport measurements are compared with results from Raman spectroscopy measurements.

---

* Email: albert.rigosi@nist.gov

# I. INTRODUCTION

Graphene continues to attract attention for its desirable electrical properties [1-4]. Though it can be synthesized in multiple ways, graphene grown epitaxially (EG) has been shown to harbor a specific set of advantages, namely its growth scale and ability to exhibit a robust quantum Hall effect [5-9]. For applications that require both of these advantages, like resistance metrology [10-11], a more restricted parameter space is immediately imposed due to the difficulty of fabricating suitable gates. For instance, EG that has been grown on large scales exhibits carrier densities that are difficult to control and has thus been subject to alternate forms of gating and doping that do not require a metallic top or bottom gate [12-14]. The application space would expand if the carrier density could be controlled over a large lateral range to fully take advantage of the types of devices into which EG may be fabricated.

In this work, the dynamics behind hole doping in EG are examined by using nitric acid as an adsorbent. Though there is some work already pertaining to this particular interaction [15-17], some elements, such as timescales associated with the desorption process, have not yet been fully understood. Further characterization of these processes is important for any two-dimensional (2D) material intended for devices with a particular functionality, such as devices exhibiting photovoltaic properties [18-19], charge density waves [20-21], or $p$-$n$ junctions [22-27]. Furthermore, the large-scale gating that becomes enabled from the understanding of these processes can also be directly applicable to photodetection [28-32] and electron optics [33-35].

Specifically, a focus is placed on quantum transport data taken at low (1.5 K) and room temperatures. A main objective for electrical measurements was to monitor the properties of a device post-exposure, with the intention of obtaining information about the timescales associated with desorption. Measurements were performed under ambient conditions to replicate the most likely situation for handling devices using this doping method. Raman spectroscopy is also employed with the main objective of comparing the relevant timescales determined from transport measurements with those from the devices' optical properties. A discussion surrounding a suitable Langmuir model supports the observations made and allows one to gain insights on how to interpret the several time constants in the data.

## II. EXPERIMENTAL AND NUMERICAL METHODS

### A. Sample Preparation

EG films were grown on 4H-SiC substrates by the high-temperature sublimation method, which allows carbon atoms on the surface to restructure into a hexagonal lattice [36]. Chips were diced from 4H-SiC(0001) wafers from CREE (see Acknowledgments), cleaned with a 5:1 diluted solution of hydrofluoric acid and deionized water, and coated with a dilute solution of the resist AZ 5214E in isopropyl alcohol to use the benefits of polymer-assisted sublimation growth (PASG) [37]. The chemically smoothed Si-face of each chip was placed on a polished glassy carbon slab (SPI Glas 22, see Acknowledgments) to limit the escape rate of the Si atoms, thereby improving graphene homogeneity. The graphite-lined resistive-element furnace (Materials Research Furnaces Inc., see Acknowledgments) was flushed with argon gas and filled to about 103 kPa from a 99.999 % liquid argon source. During growth, the furnace was held at 1900 °C for between 4 min to 5 min, with heating and cooling rates of about 1.5 °C/s.

As-grown EG films had their uniformity verified by means of confocal laser scanning microscopy (CLSM) and optical microscopy. Device fabrication closely followed steps described in other work and generally included a gold protection layer, photolithography processes, and protection layer removal [38-39]. For some devices, superconducting NbTiN was deposited as the electrical contact material as an alternate means to determine possible differences in contrast with gold contacts [40], of which none were observed. The final fabrication step for devices that did not serve as the control was to undergo a functionalization process for regulating the carrier density without the need for a top gate. The functional group $Cr(CO)_3$ was implemented in a similar manner to previous studies [13, 41].

The carrier density of these functionalized EG devices following exposure to air for about one day is on the order $10^{10}$ cm$^{-2}$ [13], and this behavior of a functionalized device to asymptotically approach the Dirac point provides a valuable comparison to the control devices, especially since the typical value of electron doping in EG can be as high as of $10^{13}$ cm$^{-2}$, in part due to the buffer layer beneath the EG layer [42-43]. A set of final device images are shown in Fig. 1 (a)-(b). First, an optical image of a finished, functionalized device is shown after wire bonding. One may notice spots within the device perimeter, which are small clusters of oxidized chromium with virtually no interaction with the EG layer [13]. A control device is shown in Fig. 1 (b), with a small blue box indicating the region where an example CLSM image was captured. This capture can be seen in Fig. 1 (d) and verifies the homogeneous growth quality of the EG. A comparison can be made with an example of an undergrown EG film in the preceding panel (Fig. 1 (c), showing material that would not be used for device fabrication).

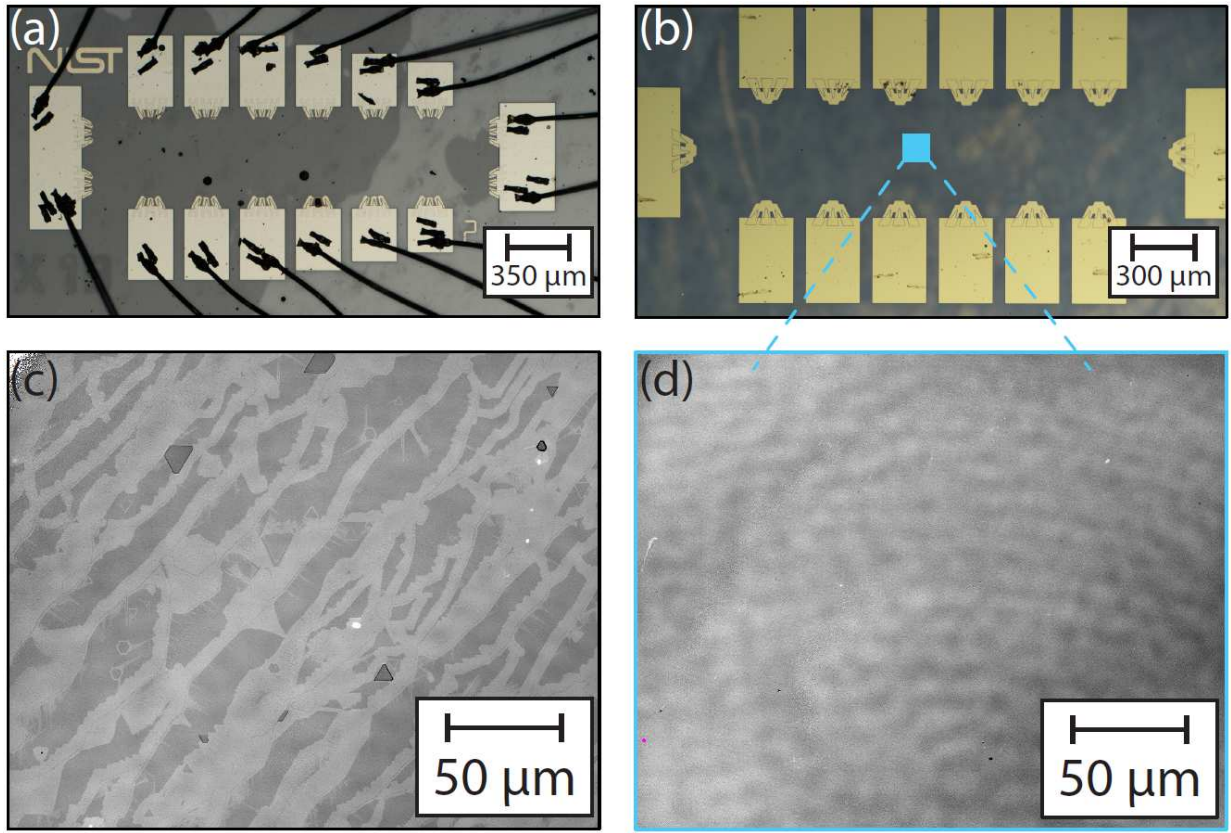

FIG. 1. (Color online) Graphene material and device quality. (a) An optical microscope image of a functionalized device is shown after wire bonding. Spots within the device perimeter are clusters of oxidized chromium that do not interact with the EG layer beneath. (b) An optical image of a standard device is shown. The small blue box indicates a region within which a closer inspection was performed to verify growth quality. (c) Confocal laser scanning microscopy was utilized to capture example images of undergrown EG on SiC to compare with (d). (d) An example CLSM image is shown to verify the homogeneous growth quality of the EG and is the relevant standard by which all grown material is compared.

## B.   Exposure and Electrical Transport

For quantum Hall transport measurements, a Janis Cryogenics system was used (see Acknowledgments). All data were collected at temperatures of 1.5 K and magnetic fields between -9 T and 9 T, primarily to determine the initial hole doping of each device. All samples were consistently exposed to nitric acid vapors in a standard fume hood, with each exposure taking place 3 cm from the surface of the liquid for a duration of 2 min. The duration was selected based on the reported faster timescales for $NO_3$ adsorption on EG [16]. Traditional lock-in measurement techniques were used for evaluating the longitudinal ($R_{xx}$) resistances for each device, and all source-drain currents were set to 1 μA.

## C. Raman Spectroscopy

Raman spectroscopy was used to both confirm the presence of a homogeneous EG layer as well as to monitor the 2D (G') mode. These measurements were performed with a Renishaw InVia micro-Raman spectrometer (see Acknowledgments) using a 633 nm wavelength excitation laser source. All spectra were measured and collected using a backscattering configuration, 1 μm spot size, 300 s acquisition time, 1.7 mW power, 50 × objective, and 1200 mm$^{-1}$ grating. In the case of confirming the quality, rectangular Raman maps were collected with step sizes of 20 μm in a 5 by 3 raster-style grid. For time-dependent Raman measurements, several sets of data were collected on single points on the film post-exposure with a 30 s delay between measurements.

## III. TRANSPORT AND TRANSIENT DOPING

When doping the control devices, an immediate transport measurement (that is, a device loaded within two minutes of exposure) was made to determine the initial hole density ($n_h$) at 1.5 K. Standard exemplary Hall measurements for a control device can be seen in Fig. 2 (a) as solid black and dashed red curves. An example set of functionalized device measurements are also shown as blue, green, and orange curves of varying dash length. The slope from each Hall curve at low magnetic fields (less than 1 T) was used for calculating $n_h$. In Fig. 2 (a), the average initial $n_h$ for the example control and functionalized devices were about $1.6 \times 10^{11}$ cm$^{-2}$ and $1.1 \times 10^{12}$ cm$^{-2}$, respectively. Another electrical measurement involves the monitoring of $R_{xx}$ after exposure to nitric acid vapor for several pairs of longitudinal contact pairs on each of the control devices. A subset of these results is shown in Fig. 2 (b), where all curves come from the same device and the same exposure.

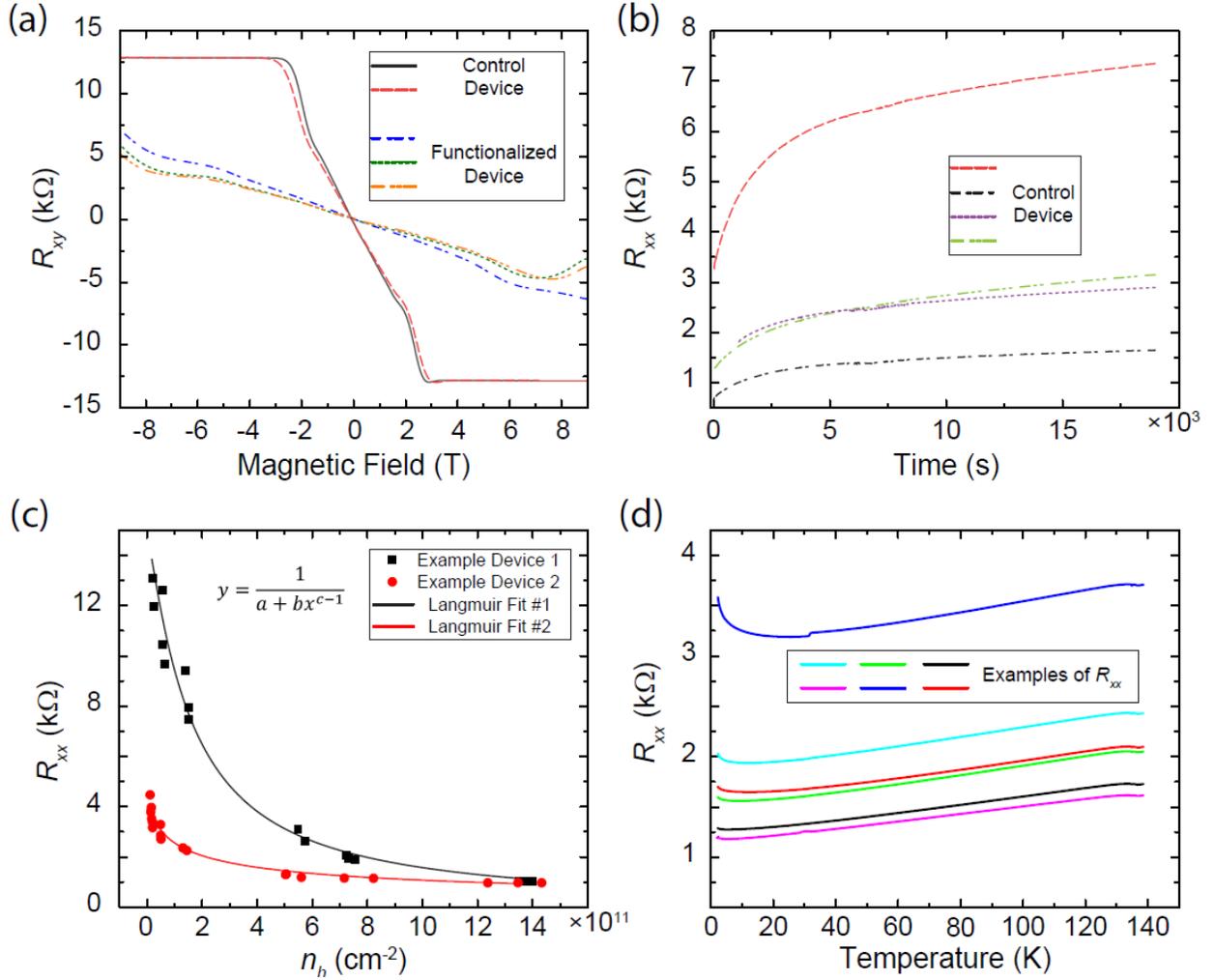

FIG. 2. (Color online) Transport and electrical data. (a) Standard Hall measurements were performed on a control device (solid black and dashed red curves) and a functionalized device (blue, green, and orange curves of varying dash length). The slopes of the Hall curves at low magnetic fields (magnitude less than 1 T) were used to calculate the initial hole densities ($n_h$)

post-exposures. (b) Longitudinal resistances ($R_{xx}$) for several pairs of longitudinal contact pairs on the control devices were measured after exposure to nitric acid. (c) For two example devices, measurements are repeated to extract an approximate relationship between $R_{xx}$ and $n_h$. These relationships are unique to each device. (d) A set of $R_{xx}$ measurements (from a single device) as a function of temperature are shown to support the notion that there is a 1:1 mapping for $R_{xx}$ values between 300 K and 1.5 K.

The observation that the nominal values of $R_{xx}$ vary significantly stems from the variation in hole density throughout the control device. This behavior is typical and one of the primary reasons that functionalization was sought as a remedy [13]. These treated devices will show more stable and predictable behavior in later figures. Ideally, one would inspect the hole doping as a function of time, as this would reveal more about the desorption processes. Therefore, for every device, low-field determinations of $n_h$ are repeated to extract an approximate relationship between $R_{xx}$ and $n_h$. For two example control devices, such relationships are plotted in Fig. 2 (c). To transform the time-dependent $R_{xx}$ curve to a time-dependent $n_h$ curve, a characteristic function and suitable fit were necessary. For reasons soon to be described, the ansatz chosen for these fits was based on a Langmuir curve (with $a$, $b$, and $c$ as constants):

$$R_{xx} = \frac{1}{a + bn_h{}^{c-1}}$$

(1)

The only remaining issue that required attention was the verification that the relationship between a room temperature $R_{xx}$ and a 1.5-K $n_h$ was reasonably a 1:1 correspondence. For instance, not meeting this criterion would imply that one resistance could have two values of $n_h$, creating inaccuracies in any time-dependent $n_h$ curve. Ergo, as shown in Fig. 2 (d), a set of $R_{xx}$ measurements, all from a single control device, are shown as a function of temperature to support the notion that there is a 1:1 mapping between $R_{xx}$ and $n_h$ values between 300 K and 1.5 K [44].

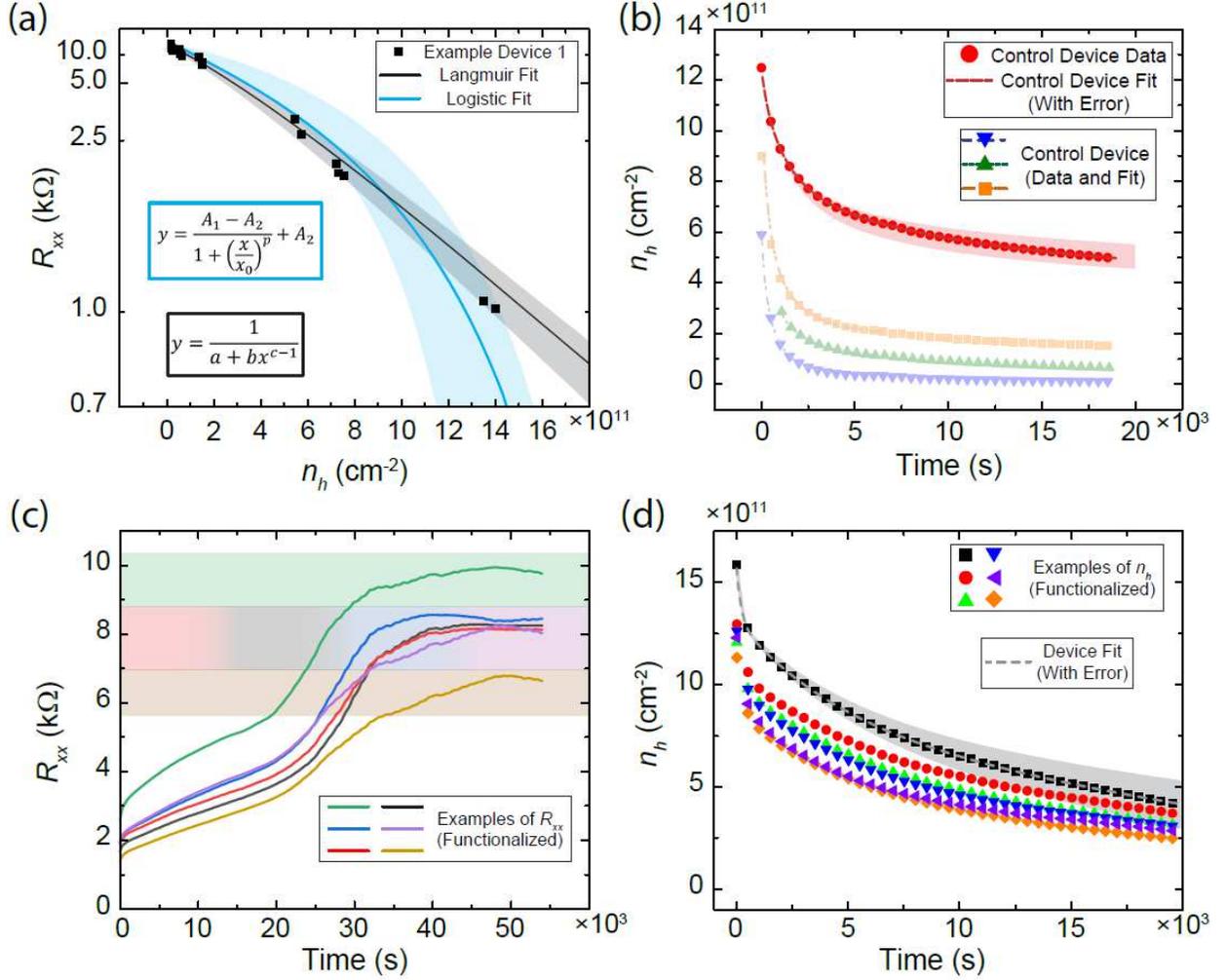

FIG. 3. (Color online) Extracting time constants. (a) An error comparison between a Langmuir fit (black) and a Logistic fit (cyan) reveals that the Langmuir fit is preferred for obtaining the relationship between $R_{xx}$ and $n_h$. The blue and gray shaded region indicate 1σ error from the fitting procedure. (b) Each Langmuir fit is used to transform the corresponding $R_{xx}$ data to $n_h$. The shaded region indicates 1σ error from the fitting procedure and is shown only for the top data (red circles) to prevent significant overlap for the other three curves. (c) Example $R_{xx}$ data from functionalized device are shown over a longer duration to emphasize the stabilizing nature of the treatment. Shaded regions are color-matched to the corresponding curves and indicate a region within which the polarity of the charge carrier may change due to the proximity of the Fermi level to the Dirac point. (d) The same data in (c) are transformed to provide a way for extracting time constants that characterize certain doping behaviors. One example fit (and corresponding 1σ error as a shaded region) is shown for the topmost curve.

To justify the use of the Langmuir fit, one may start with the fact that the ideal relationship between the two quantities $R_{xx}$ and $n_h$ is approximately reciprocal [45]. One additional possibility for fitting the data (shown in Fig. 3 (a)) is to use the Logistic fit below (where $A_1$, $A_2$, $x_0$, and $p$ are constants):

$$R_{xx} = \frac{A_1 - A_2}{1 + \left(\frac{n_h}{x_0}\right)^p} + A_2$$

$$(2)$$

When considering the errors associated with the fitting procedure (performed with OriginLab, see Acknowledgments), one can compare the adequacy of the fits, as done in Fig. 3 (a) for an example control device. A Langmuir fit (black) and a Logistic fit (cyan) are compared, and from a visual analysis, it became clear that the Langmuir fit would be a better indicator of the relationship between $R_{xx}$ and $n_h$. Furthermore, an optimal reduced chi-squared validated this determination. The blue and gray shaded region indicate 1σ error from the fitting procedure. The obtained Langmuir fits (as shown in Fig. 3 (b)) were used to transform the corresponding $R_{xx}$ data to $n_h$. The shaded region indicates 1σ error from the fitting procedure and is shown only for the top data (red circles) to prevent significant overlap for the other three curves. These example curves correspond to Fig. 2 (b).

A similar transformation can be performed for data acquired with functionalized devices. For instance, an example set of $R_{xx}$ data are shown in Fig. 3 (c) over a longer duration to emphasize the asymptotic behavior induced by the treatment. The shaded regions are color-matched to their corresponding curves to show the approximate range of longitudinal resistances within which the polarity of the charge carrier is expected to change indiscriminately due to the proximity of the Fermi level to the Dirac point. The final transformed example data are then shown in Fig. 3 (d) to demonstrate both that functionalized devices have a smaller variation of hole density across the device and that similar time constants can be extracted. One example fit is shown in dashed gray (with a shaded gray region corresponding to a 1σ fitting error). An additional note to make is that functionalized devices consistently started at the same initial $n_h$ after exposure (to within parts in $10^{10}$ cm$^{-2}$), providing a more predictable doping process than their control counterparts, with an initial $n_h$ varying by, at most, a few parts in $10^{11}$ cm$^{-2}$.

All the data in Fig. 3 (d) were then fitted with three-term exponential decays to account for several expected desorbed species of molecule (*i.e.* NO$_3$, NO$_2$, and water from potential dissociation of HNO$_3$ that has physisorbed on the EG surface) [16]. Three-term decays provided an optimized reduced chi-squared when compared to double- or single-term decays. The three time constants ($\tau_1$, $\tau_2$, and $\tau_3$) averaged from all the fits corresponding to the data are 204 s ± 95 s, 2575 s ± 326 s, and 7.35 × 10$^4$ s ± 2.43 × 10$^4$ s, respectively. To fully understand these constants, more analysis and discussion on Langmuir modeling is required (not to be confused with the previously described Langmuir fit needed for the transformation between $R_{xx}$ and $n_h$).

## IV. LANGMUIR MODELING AND RAMAN MONITORING

### A. Three-Species Langmuir Model

The interpretation of these time constants required further analysis. If it is assumed that adsorbents have saturated the EG surface after the exposure, this analysis becomes more straightforward for short monitoring times. Some work discussing $NO_2$ adsorption on EG show that the associated time constants for this molecule are on the order of 100 s [46-47]. This description fits well within the extracted value of $\tau_1$. Combined with the reports that the adsorption of oxygen and water on graphene takes place on the order of 2 h [48-52], there should not be any immediate competing effects at such short timescales. The longest timescale ($\tau_3$) may be compared with a very similar value obtained from a previous work that described the time constant as representing the desorption of water from the EG surface (that value was reported to be approximately $7 \times 10^4$ s) [13]. This leaves the remaining component of the nitric acid byproducts, $NO_3$, to be described by $\tau_2$.

Though it was shown that water was the dominant contributor to desorption when competing with oxygen [13], it cannot be immediately assumed for a three species competition at longer times (long enough to neglect contributions from $NO_2$). Though a two species Langmuir model has been shown in the literature [53], a model for three competing species has not. To develop this model, three ordinary differential equations were set up that describe the occupancies of oxygen, water, and $NO_3$ on EG sites [53]:

$$\frac{d\theta_{i,j,k}}{dt} = k_{i,j,k(A)} p_{i,j,k} \left(1 - \theta_i(t) - \theta_j(t) - \theta_k(t)\right) - k_{i,j,k(D)} \theta_{i,j,k}(t)$$

(3)

Here, $i$, $j$, and $k$ represent each of the three competing species. In all cases, $k_{(A)}$ and $k_{(D)}$ are adsorption and desorption rate constants, respectively. Rate constants take on the form: $k_{A,D} = \nu \, exp\left[\frac{-E_{ads,des}}{k_B T}\right]$, with $E_{ads}$ and $E_{des}$ as the adsorption and desorption energy per molecule, respectively, $\nu$ is the attempt frequency (typically approximated as $10^{13}$ s$^{-1}$ [13]), $k_B$ is the Boltzmann constant, and $T$ is temperature (K). The term $P_{(LD)}$ denotes the Langmuir desorption pressure, unique to each of the gas species, with the form:

$$P_{LD} = \frac{k_B T}{\left(\frac{h^2}{2\pi m k_B T}\right)^{3/2}}$$

(4)

The listed quantities in all equations should be converted to SI units (or remain unitless) to avoid unit confusion within the expressions. In Eq. 4, $h$ is the Planck constant, and $m$ is the mass of a single molecule of one of the gas species (kg). Combining all of these elements with Eq. 3 allows one to solve for equilibrium occupancies, where the steady state solutions may be written as [53]:

$$\theta_{i,j,k} = \frac{p_{i,j,k}}{p_{i,j,k} + P_{i,j,k\,(LD)}\,exp\left[\frac{-E_{i,j,k}}{k_B T}\right]\left[1 + \frac{p_{j,k,i}}{P_{j,k,i\,(LD)}}\,exp\left[\frac{E_{j,k,i}}{k_B T}\right] + \frac{p_{k,i,j}}{P_{k,i,j\,(LD)}}\,exp\left[\frac{E_{k,i,j}}{k_B T}\right]\right]}$$

(5)

In Eq. 5, $p$ is the partial pressure (with the fractional form multiplied by the total pressure) of the gas species, which is about 0.209 for oxygen and 0.00916 for water at room temperature and 40 % relative humidity under normal ambient conditions. The net adsorption energy $E$ is 0.15 eV and 0.1 eV for oxygen and water, respectively [49-50]. An estimate of the partial pressure for $NO_3$ is needed to obtain occupancy information. This can be done using Fick's laws of diffusion along with the conditions that there are no significant air currents (as is the case for the used probe, whose sample holder resides within a metallic encasement). The mean squared displacement may be estimated as $6Dt$, where $t$ is the time spent diffusing and $D$ is the diffusion coefficient, which may be approximated as $10^{-5}$ m$^2$/s [54]. The condition is that a saturated sample may have most of its sites occupied, and for a 1 cm$^2$ area, this would amount to about $10^{15}$ cm$^{-2}$ molecules.

With these conditions, the square root of the mean squared displacement can be calculated to get approximate distances of 1 cm, 25 cm, and 75 cm for the corresponding example times of $t_\alpha$ = 200 s, $t_\beta$ = 1000 s, and $t_\gamma$ = 8000 s, respectively, after the exposure. These times were selected to avoid any dominant transient effects from NO$_2$. Spreading these molecules out from the surface of EG to the volume of diffusion yields the following three unitless partial pressures: $p_\alpha = 10^{-4}$, $p_\beta = 4 \times 10^{-6}$, and $p_\gamma = 1.33 \times 10^{-6}$. The final element needed is net adsorption energy, which can be approximated by the rate constant formula after Eq. 3. By inverting $\tau_{NO_3}$ ($E_{des} \approx 1$ eV) and the adsorption saturation time of 10 s (upper bound, $E_{ads} \approx 0.8$ eV) from Ref. [16], the net 0.2 eV allows one to calculate occupancies. Note that in Ref. [13], the calculation for two competing species yielded the result of about $\theta_{O_2} = 89.3$ % and $\theta_{H_2O} = 10.7$ %.

If it is assumed that the partial pressure at each time is held constant, then the following occupancies serve as a lower bound estimate given the saturation of NO$_3$. At 200 s, the expected steady state occupancies are: $\theta_{O_2} = 84.4$ %, $\theta_{O_2} = 10.1$ %, and $\theta_{H_2O} = 5.5$ %. At 1000 s, the resulting occupancies are: $\theta_{O_2} = 89.0$ %, $\theta_{O_2} = 10.7$ %, and $\theta_{H_2O} = 0.3$ %. Lastly, at 8000 s, the results are: $\theta_{O_2} = 89.2$ %, $\theta_{O_2} = 10.7$ %, and $\theta_{H_2O} = 0.1$ %. This behavior is consistent with the case of two competing species since NO$_3$ is not a significant atmospheric constituent. This analysis also shows that water remains a dominant desorbing agent for longer timescales, giving additional confirmation that $\tau_{NO_3} \approx 2575$ s $\pm$ 326 s under standard atmospheric conditions.

## B. Monitoring the 2D (G') Raman Mode

Additional evidence of the inherent timescales associated with this transient hole doping was collected via Raman spectroscopy performed on functionalized devices. After collecting a time series of Raman spectra, examples of which are shown in Fig. 4 (a), the 2D (G') mode frequencies required analysis. The four example spectra were observed at four distinct times labeled $t_1$ through $t_4$ and correspond to 0 s, $5 \times 10^3$ s, $10^4$ s, and $1.2 \times 10^4$ s, respectively. The peaks at all times were fit to a Lorentzian profile to extract the frequency. The fitting results are shown in Fig. 4 (c) as black data points with error bars indicating 1σ error from the peak fitting procedure. The issue now becomes how to determine the relationship between the 2D mode frequency and $n_h$, and two analyses were conducted to understand this relationship.

The first analysis involves using the existing literature to calculate an expected behavior of the 2D mode frequency with time. To accomplish this, a best-fit cubic curve was used (extracted from Ref. [55], shown by the inset of Fig. 4 (b)). This curve relating Fermi energy and 2D mode frequency was recalculated to instead show the relationship between the 2D mode frequency and $n_h$ via $E_F = \hbar v_F \sqrt{\pi |n_h|} sign(n_h)$. This recalculation used two known conditions: (1) $n_h$ was determined by electrical transport and was predictable (about $1.1 \times 10^{12}$ cm$^{-2}$), and (2) $n_h$ could be approximated at the minimum wavenumber as corresponding to the Dirac point given the results of Fig. 3 (c). Using this recalculation, average time-dependent $n_h$ curves were transformed to calculate a predicted Raman peak shifting (gold) based on $\tau_{NO_3}$ in Fig. 4 (c). The dashed cyan curves are repeated calculations based on different time constants to show the extent of the prediction accuracy.

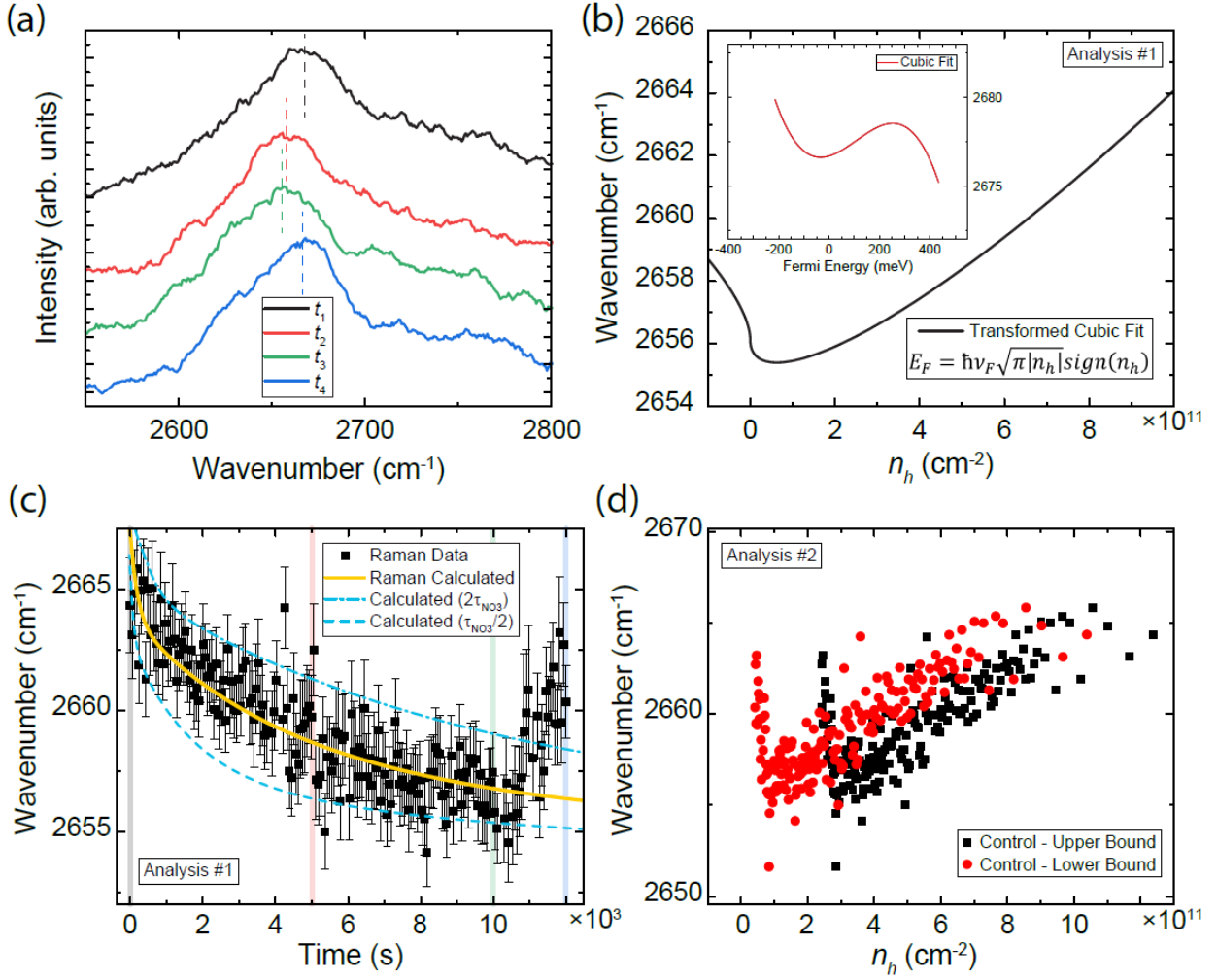

FIG. 4. (Color online) Spectroscopic verification of observed time constants. (a) Example Raman spectra focused on the 2D (G′) mode of EG. These four spectra are observed at four distinct times labeled $t_1$ through $t_4$ and correspond to 0 s, $5 \times 10^3$ s, $10^4$ s, and $1.2 \times 10^4$ s, respectively. (b) The relationship between the 2D mode frequency and $n_h$ is calculated by using a best-fit cubic curve as extracted from Ref. [55] (inset, which notes the 2D mode frequency as a function of Fermi energy). For this first analysis, this transformation is given two initial conditions: the first $n_h$ is known and $n_h$ is known at the minimum wavenumber (corresponding to the Dirac point). (c) The 2D mode frequencies for the time series measurements are plotted (with error bars indicating 1σ error from the peak fitting procedure). Calculated trends of the Raman peak shifting are plotted in gold (solid) and cyan (dashed) based on the relevant time constant. Four thin, color-matched bands mark the same times listed in (a). (d) A second analysis reveals a relationship between the 2D mode frequency and $n_h$ as extracted from the data directly, with the mode's doping-dependence showing reasonable behavior when compared to the literature.

It should be noted that the prediction of the Raman trend (gold curve) is a simple transformation that gives a 1:1 correspondence between 2D mode frequency and $n_h$. The immediate rise in wavenumber that occurs after the minimum can be attributed to the change in polarity of the charge carrier. It behaves reasonably considering that a similar trend can be seen in Fig. 4 (b) as one considers crossing the Dirac point to obtain an electron density (or below $n_h \approx 0$ cm$^{-2}$).

For the second analysis, one may decide to forego the use of predictive assistance from the literature. The alternate approach would be to use the time-dependent $n_h$ curve as a direct method to reveal the relationship between the 2D mode frequency and $n_h$. On the one hand, the mode's doping-dependence, shown in Fig. 4 (d) for an example control device, shows a reasonable behavior when compared to the literature [55-56]. The major issue that arises from this alternate approach is that the wavenumber extrema appear to occur at some arbitrary positive value of $n_h$. This horizontal departure from the calculation in Fig. 4 (b) demonstrates that, especially for devices with more variation in $n_h$, an unexplainable minimum can arise in the final transformed data. That said, some of the error could be accounted for by means of introducing a rigorous horizontal error bar, but this ultimately reduces the predictive quality between a device's hole doping and the frequency of its 2D mode.

## V. CONCLUSIONS

In this work, the dynamics of transient hole doping are reported for epitaxial graphene devices that are both untreated and functionalized. Nitric acid was used as the adsorbent, and corresponding timescales associated with its desorption were determined from transport and electrical monitoring data. The understanding of reversible hole doping without gating is of crucial importance to those seeking to fabricate devices on length scales where metallic gating becomes unfeasible. Optical properties were also monitored with time after exposure via Raman spectroscopy, supporting the determined time constants for $NO_3$. These results will be relevant for future device fabrication involving any material that can be hole (or electron) doped by chemical means.

## ACKNOWLEDGMENTS AND NOTES

The authors thank L. S. Chao, A. L. Levy, G. J. Fitzpatrick, and E. C. Benck for their assistance with the NIST internal review process. Work presented herein was performed, for a subset of the authors, as part of their official duties for the United States Government. Funding is hence appropriated by the United States Congress directly. The authors declare no competing interest.

Commercial equipment, instruments, and materials are identified in this paper in order to specify the experimental procedure adequately. Such identification is not intended to imply recommendation or endorsement by the National Institute of Standards and Technology or the United States Government, nor is it intended to imply that the materials or equipment identified are necessarily the best available for the purpose.